\documentclass[useAMS,usegraphicx,usenatbib]{mn2e}

\newcommand{\aap}{A\&A}
\newcommand{\apj}{ApJ}
\newcommand{\apjl}{ApJ}
\newcommand{\apjs}{ApJS}
\newcommand{\araa}{ARA\&A}
\newcommand{\mnras}{MNRAS}
\newcommand{\nat}{Nat}
\newcommand{\physrep}{Phys. Rep.}

\title[The First Galaxies]{The first galaxies: assembly, cooling and the onset of turbulence}

\author[Greif et al.]{Thomas H. Greif$^{1,2,3}$\thanks{E-mail: tgreif@ita.uni-heidelberg.de}, Jarrett L. Johnson$^{2}$, Ralf S. Klessen$^{1}$ and Volker Bromm$^{2}$\\$^{1}$ Institut f\"{u}r theoretische Astrophysik, Albert-Ueberle Strasse 2, 69120 Heidelberg, Germany\\$^{2}$ Department of Astronomy, University of Texas, Austin, TX 78712, USA\\$^{3}$ Fellow of the International Max Planck Research School for Astronomy and Cosmic Physics at the University of Heidelberg}

\begin{document}

\maketitle
\topmargin-1cm

\begin{abstract}
We investigate the properties of the first galaxies at $z\ga 10$ with highly resolved numerical simulations, starting from cosmological initial conditions and taking into account all relevant primordial chemistry and cooling. A first galaxy is characterized by the onset of atomic hydrogen cooling, once the virial temperature exceeds $\simeq 10^{4}~\rm{K}$, and its ability to retain photoheated gas. We follow the complex accretion and star formation history of a $\simeq 5\times 10^{7}~\rm{M}_{\odot}$ system by means of a detailed merger tree and derive an upper limit on the number of Population~III (Pop~III) stars formed prior to its assembly. We investigate the thermal and chemical evolution of infalling gas and find that partial ionization at temperatures $\ga 10^{4}~\rm{K}$ catalyses the formation of H$_{2}$ and hydrogen deuteride, allowing the gas to cool to the temperature of the cosmic microwave background. Depending on the strength of radiative and chemical feedback, primordial star formation might be dominated by intermediate-mass Pop~III stars formed during the assembly of the first galaxies. Accretion on to the nascent galaxy begins with hot accretion, where gas is accreted directly from the intergalactic medium and shock-heated to the virial temperature, but is quickly accompanied by a phase of cold accretion, where the gas cools in filaments before flowing into the parent halo with high velocities. The latter drives supersonic turbulence at the centre of the galaxy and could lead to very efficient chemical mixing. The onset of turbulence in the first galaxies thus likely marks the transition to Pop~II star formation.
\end{abstract}


\section{Introduction}
Understanding the formation of the first stars and galaxies at the end of the cosmic dark ages is one of the most important challenges in modern cosmology \citep{bl01,bl04a,cf05,glover05}. In the standard $\Lambda$ cold dark matter ($\Lambda\rm{CDM}$) cosmology, the first stars, termed Population~III (Pop~III), are predicted to have formed in dark matter (DM) `minihaloes' with virial mass $\sim 10^{6}~\rm{M}_{\odot}$ at $z\sim 20$ \citep{bcl99,bcl02,nu01,abn00,abn02,yoshida03a,yoshida06b,gao07,on07}. Based on molecular hydrogen cooling, the gas can cool to $\simeq 200~\rm{K}$ and become Jeans-unstable once the central clump attains $\simeq 10^{3}~\rm{M}_{\odot}$. Such high temperatures lead to efficient accretion on to the protostellar core and imply that the first stars were predominantly very massive, with $M_{*}\ga 100~\rm{M}_{\odot}$ (\citealt{op03,bl04b,on07}; but see \citealt{tm04,ripamonti07,mt07,mb08}).

Due to their primordial composition, massive Pop~III stars have smaller radii than their present-day counterparts and their surface temperatures can exceed $\simeq 10^{5}~\rm{K}$, resulting in high photon yields \citep{ts00,bkl01,schaerer02,tsv03}. Consequently, they exert strong feedback on the intergalactic medium (IGM) via radiation in the Lyman--Werner (LW) bands, which readily destroys molecular hydrogen \citep{gb01,mba01,rgs01,oy03,gb06,mbh06,on08,jgb07b,wa07a}, and ionizing radiation giving rise to the first H~{\sc ii} regions \citep{kitayama04,wan04,abs06,su06,yoshida07,as07,awb07,jgb07a,whalen07}. Furthermore, a yet unknown fraction of massive, metal-free stars is expected to end in extremely violent supernova (SN) explosions \citep{hw02,heger03}, profoundly affecting the IGM in terms of dynamics and chemical enrichment \citep{byh03,mbh03,nop04,sfs04,ybh04,ky05,machida05,greif07,whalen08}. Since primordial star formation initially occurs in minihaloes that constitute the progenitors of the first galaxies, stellar feedback is expected to play an important role during their assembly.

When and where did the first galaxies form? According to the bottom-up character of structure formation, as described by standard Press-Schechter theory \citep{ps74}, the first $\ga 5\times 10^{7}~\rm{M}_{\odot}$ haloes assembled at $z\ga 10$ via hierarchical merging. This mass scale is set by the onset of atomic hydrogen cooling once the virial temperature exceeds $\simeq 10^{4}~\rm{K}$; these objects have therefore frequently been termed `atomic cooling haloes'. Henceforth, we will synonymously use the term `first galaxy', as these haloes can retain photoheated gas and therefore might, for the first time, maintain self-regulated star formation in a multi-phase, interstellar medium \citep{tw96,mf99,mfr01,mfm02,oh02,rgs02a,rgs02b,sfm02,tsd02,wv03,dijkstra04,rpv06,rgs08}.

What types of stars are expected to form during the assembly of the first galaxies? Depending on the primordial initial mass function (IMF) and the strength of radiative and SN-driven feedback, the IGM could be heavily enriched with metals prior to the onset of second-generation star formation \citep{byh03,greif07,wa07c}. The existence and precise value of a critical metallicity for the transition to Pop~II are still vigorously debated, and has originally been discussed in the context of fine-structure versus dust cooling \citep{bl03a,schneider03,fjb07,jappsen07a,ss07}. In the former case, critical metallicities of $\rm{Z_{crit}} \simeq 10^{-3.5}~\rm{Z}_{\odot}$ have been found \citep{bromm01,ss06a}, while in the latter case uncertainties in the dust composition and gas-phase depletion lead to a range of possible values, $10^{-6}\la \rm{Z_{crit}}\la 10^{-5}~\rm{Z}_{\odot}$ \citep{omukai05,schneider06,to06}. Somewhat transcending this debate, recent simulations have shown that a single SN explosion might enrich the local IGM to well above any critical metallicity \citep{byh03,greif07,wa07c}, forcing an early transition in star formation mode \citep{cgk08}. However, if radiative feedback was sufficiently strong, the bulk of primordial star formation might have occurred in systems that cool via atomic hydrogen lines \citep{gb06,on08,jgb07b}. In light of these uncertainties, we investigate the formation of a first galaxy in the limiting case of no feedback, allowing us to focus on the chemistry, cooling and development of turbulence during the assembly of the atomic cooling halo.

Theoretical investigations have recently pointed towards the existence of two physically distinct populations of metal-free stars \citep{ui00,mbh03,jb06}. Gas cooling primarily via molecular hydrogen leads to the formation of $\ga 100~\rm{M}_{\odot}$ stars, termed Pop~III.1, while in regions of previous ionization hydrogen deuteride (HD) likely enables the formation of $\ga 10~\rm{M}_{\odot}$ stars, termed Pop~III.2 \citep{tm08}. This latter mode had previously been termed Pop II.5 \citep{mbh03,gb06,jb06}. Scenarios providing a substantial degree of ionization include relic H~{\sc ii} regions \citep{no05}, dense shells produced by energetic SN explosions \citep{byh03,mbh03,sfs04,machida05}, and structure formation shocks in the virialization of the first galaxies \citep{oh02,gb06}. In the former case, numerical simulations have confirmed that the gas can cool to the temperature of the cosmic microwave background (CMB), reducing the Jeans-mass by almost an order of magnitude compared to the truly primordial case \citep{jgb07a,yoh07}, while we here set out to investigate the effects of partial ionization and molecule formation during the assembly of the first galaxies. One of the key questions is therefore whether the gas can cool to the CMB limit and thus enable the formation of Pop~III.2 stars.

Another important aspect concerning the baryonic collapse of the first galaxies is the development of turbulence. While turbulence does not seem to play a role in minihaloes, at least on scales comparable to their virial radius \citep{yoshida06b}, turbulent motions could become important in more massive haloes, where accretion of cold gas from filaments in combination with a softened equation of state drives strong shocks \citep{keres05,wa07b,sancisi08}, possibly leading to vigorous fragmentation and the formation of the first star clusters \citep{cgk08}. This would mark the first step towards conditions similar to present-day star formation, where supersonic turbulent velocity fields determine the fragmentation properties of the gas. In our present study, we investigate the role of turbulence by analysing the velocity field and energy content of the galaxy during virialization. We then discuss the likely fragmentation properties of the gas and the consequences for second-generation star formation.

The final ingredient of early galaxy formation is the co-evolution of massive black holes (MBHs) and the surrounding stellar system, leading in extreme cases to the formation of the highest redshift quasars observed at $z\ga 6$. The most sophisticated simulations of supermassive black hole (SMBH) growth to date have been conducted by \citet{li07}, but their resolution was insufficient to trace the origin of SMBHs below $\simeq 10^{5}~\rm{M}_{\odot}$. Such SMBHs could form via direct collapse of isothermal gas in atomic cooling haloes in the presence of a strong photo-dissociating background \citep{bl03b,kbd04,bvr06,ln06,ss06b,wta07}, or via merging and accretion of gas on to the compact remnants of Pop~III stars \citep{mr01,schneider02,its03,vhm03,vr05,jb07,pdc07,wta07}. Here, we investigate the fate of BHs seeded by the remnants of Pop~III stars and their prospect of growing into SMBHs at the centres of the first galaxies. Finally, we provide an estimate for the amplitude of ionizing and molecule-dissociating radiation emitted by the accretion disc around the central BH.

The structure of our work is as follows. In Section~2, we discuss the setup of the cosmological simulations, the implementation of multiple levels of refinement and our sink particle algorithm. We then describe the hierarchical assembly of the galaxy (Section~3), its cooling and star formation properties (Section~4) and the development of turbulence (Section~5). In Section~6, we discuss the growth of the BH at the centre of the galaxy, and in Section~7 we summarize our results and assess their implications. For consistency, all distances are physical (proper), unless noted otherwise.

\section{Numerical methodology}
To investigate the formation of the first galaxies with adequate accuracy, our simulations must resolve substructure on small scales as well as tidal torques and global gravitational collapse on much larger scales. We capture all the relevant dynamics by performing a preliminary simulation with a coarse base resolution, but refine around the point of highest fluctuation power. This ensures that the region containing the mass of the galaxy is well resolved and that a given number of particles is distributed efficiently.

\subsection{Initial setup}
The simulations are carried out in a cosmological box of linear size $\simeq 700~\rm{kpc}$ (comoving), and are initialized at $z=99$ according to a concordance $\Lambda$ cold dark matter ($\Lambda$CDM) cosmology with matter density $\Omega_{\rm{m}}=1-\Omega_{\Lambda}=0.3$, baryon density $\Omega_{\rm{b}}=0.04$, Hubble parameter $h=H_{0}/\left(100~\rm{km}~\rm{s}^{-1}~\rm{Mpc}^{-1}\right)=0.7$, spectral index $n_{\rm{s}}=1.0$, and normalization $\sigma_{8}=0.9$ \citep{spergel03}. Density and velocity perturbations are imprinted at recombination with a Gaussian distribution, and we apply the Zeldovich approximation to propagate the fluctuations to $z=99$, when the simulation is started. To capture the chemical evolution of the gas, we follow the abundances of H, H$^{+}$, H$^{-}$, H$_{2}$, H$_{2}^{+}$, He, He$^{+}$, He$^{++}$, and e$^{-}$, as well as the five deuterium species D, D$^{+}$, D$^{-}$, HD and HD$^{+}$.  We include all relevant cooling mechanisms, i.e. H and He atomic line cooling, bremsstrahlung, inverse Compton scattering, and collisional excitation cooling via H$_{2}$ and HD \citep[see][]{jb06}.

\subsection{Refinement and sink particle formation}
In a preliminary run with $64^3$ particles per species (DM and gas), we locate the formation site of the first $\simeq 5\times 10^{7}~\rm{M}_{\odot}$ halo. This object is just massive enough to activate atomic hydrogen cooling and fulfil our prescription for a galaxy. We subsequently carry out a standard hierarchical zoom-in procedure to achieve high mass resolution inside the region destined to collapse into the galaxy \citep[e.g.][]{nw94,tbw97,gao05}. We apply three consecutive levels of refinement centred on this location, such that a single parent particle is replaced by a maximum of $512$ child particles. The particle mass in the region containing the comoving volume of the galaxy with an extent of $\simeq 200~\rm{kpc}$ (comoving) is $\simeq 100~\rm{M}_{\odot}$ in DM and $\simeq 10~\rm{M}_{\odot}$ in gas. This allows a baryonic resolution of $\simeq 10^{3}~\rm{M}_{\odot}$, such that the `loitering state' in minihaloes at $T\simeq 200~\rm{K}$ and $n_{\rm{H}}\simeq 10^{4}~\rm{cm}^{-3}$ is marginally resolved \citep{bcl02}.

To follow the growth of BHs seeded by the collapse of Pop~III.1 stars in minihaloes, we apply a slightly modified version of the sink particle algorithm introduced in \citet{jb07}. Specifically, we create sink particles once the density exceeds $n_{\rm{H}}=10^{4}~\rm{cm}^{-3}$ and immediately accrete all particles within the resolution limit of the highest-density particle, resulting in an initial sink particle mass of $\simeq 2\times 10^{3}~\rm{M}_{\odot}$. Further accretion is governed by the following criteria: particles must be bound ($E_{\rm{kin}}<\left|E_{\rm{pot}}\right|$), have a negative divergence in velocity and fall below the Bondi radius:
\begin{equation}
r_{\rm{B}}=\frac{\mu m_{\rm{H}}GM_{\rm{BH}}}{k_{\rm{B}}T}\mbox{\ ,}
\end{equation}
where $\mu$ is the mean molecular weight, $M_{\rm{BH}}$ the mass of the black hole, and $T$ the temperature of the gas at the location of the sink particle, determined by a mass-weighted average over all accreted particles. Typical values for the initial Bondi radius are $r_{\rm B}\simeq 5~\rm{pc}$. Here, we do not model radiative feedback that would result from accretion, and we therefore overestimate the Bondi radius and consequently the growth of the BH. We further assume that BH mergers occur instantaneously once their separation falls below the resolution limit, provided that their relative velocity is smaller than the escape velocity. Our idealized treatment provides a robust upper limit on the growth rate of BHs during the assembly of the first galaxies, but a more realistic calculation including the effects of feedback will be presented in future work.

\section{First galaxy assembly}
Among the key questions concerning the formation of the first galaxies are the degree of complexity associated with the halo assembly process, the role of previous star formation in minihaloes, the chemical and thermal evolution of infalling gas, and the development of turbulence. To answer these questions, we first discuss the gravitational evolution of the DM with a merger tree that reconstructs the mass accretion history of the resulting atomic cooling halo, and allows us to determine the maximum amount of previous star formation in its progenitor haloes.

\subsection{Atomic cooling criterion}
Possibly the most important characteristic distinguishing the first galaxies from their lower-mass minihalo predecessors is their ability to cool via atomic hydrogen lines, which softens the equation of state at the virial radius and allows a fraction of the potential energy to be converted into kinetic energy. At high redshift, the virial temperature of a system with virial mass $M_{\rm{vir}}$ can be expressed as
\begin{equation}
T_{\rm{vir}}\simeq 10^{4}\,\rm{K\,}\left(\frac{M_{\rm{vir}}}{5\times 10^{7}~\rm{M}_{\odot}}\right)^{2/3}\left(\frac{1+z}{10}\right)\mbox{\ ,}
\end{equation}
such that a $5\times 10^{7}~\rm{M}_{\odot}$ halo at $z\simeq 10$ is just massive enough to fulfil the atomic cooling criterion \citep{oh02}. Related to this, they have the ability to retain gas photoheated by hydrogen ionization \citep{dijkstra04}, likely allowing self-regulated star formation for the first time. Furthermore, it is often argued that star formation in atomic cooling haloes provided the bulk of the photons for reionization due to efficient shielding from LW radiation \citep{bl03b,gb06,hb06,cfg08,wa07d}. They may have been the key drivers of early IGM enrichment \citep{mfr01}, and the host systems for the formation of the first low-mass stars that can be probed in the present-day Milky Way via stellar archaeology \citep{fjb07,kjb07}. Their formation thus marks an important milestone in cosmic history, and in light of upcoming observations it is particularly important to understand the properties of these systems.

\subsection{Cosmological abundance}
How common are atomic cooling haloes at the redshifts we consider? A frequently used tool to estimate their abundance is the Press-Schechter mass function \citep{ps74}, which provides an analytic expression for the number of DM haloes per mass and comoving volume. Even though the Press-Schechter mass function is inaccurate at low redshifts compared to other analytic estimates such as the Sheth-Tormen mass function \citep{st02}, it provides a good fit to numerical simulations at the early times we are interested in \citep{jh01,heitmann06,reed07}. In Fig.~1, we show the Press-Schechter mass functions for $z\simeq 30$ and $10$ (solid and dotted line, respectively), while the symbols indicate the halo distribution extracted from the simulation by the group-finding algorithm HOP \citep{eh98}. The simulation and the analytic estimate generally agree, although there is a large scatter due to the finite box size. Better agreement could be expected if one accounted for the fluctuation power on scales greater than the computational box \citep[see][]{yoshida03b}. At redshift $z\simeq 10$, one expects to find roughly $10$ atomic cooling haloes per cubic $\rm{Mpc}$ (comoving), i.e. of the order of unity in our computational box of length $\simeq 700~\rm{kpc}$ (comoving). Indeed, in our simulation a single $5\times 10^{7}~\rm{M}_{\odot}$ halo forms by $z\simeq 10$.

\begin{figure}
\begin{center}
\includegraphics[width=8cm]{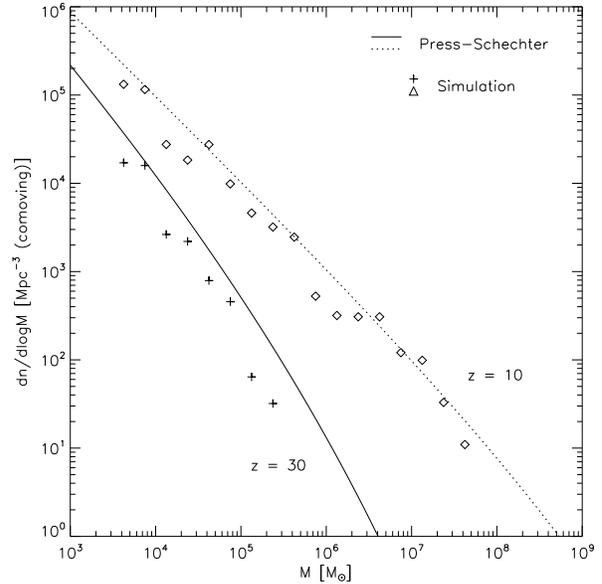}
\caption{Comparison of the Press-Schechter mass function (solid and dotted line) and the simulation results (crosses and triangles) at $z\simeq 30$ (lower set) and $z\simeq 10$ (upper set). The simulation generally agrees with the analytic prediction, although there is a large scatter due to the finite box size. The expected number of atomic cooling haloes per cubic $\rm{Mpc}$ (comoving) at $z\simeq 10$ is of the order of $10$.}
\end{center}
\end{figure}

\subsection{Assembly of atomic cooling halo}
In the $\Lambda$CDM paradigm, structure formation proceeds hierarchically, with small objects collapsing first and subsequently growing via merging. This behaviour eventually leads to the formation of haloes massive enough to fulfil the atomic cooling criterion. In Fig.~2, we show the DM overdensity, gas density and temperature averaged along the line of sight at three different output times. The brightest regions in the DM distribution mark haloes in virial equilibrium, according to the commonly used criterion $\rho/\bar{\rho}\simeq 178$, where $\rho$ and $\bar{\rho}$ denote the local and background density. White crosses denote the formation sites of Pop~III.1 stars in minihaloes. The first star-forming minihalo at the centre of the box assembles at $z\simeq 23$ and subsequently grows into the galaxy delineated by the insets in the right panels of Fig.~2 and further enlarged in Figs~8, 9, 10 and 12. Although this structure is not yet fully virialized and exhibits a number of sub-components, it has a common potential well and attracts gas from the IGM towards its centre of mass, where it is accreted by the central BH once it falls below the Bondi radius. The virial temperature of the first minihalo increases according to equation~(2) until it reaches $\simeq 10^{4}~\rm{K}$ and atomic cooling sets in, at which point the equation of state softens and a fraction of the potential energy is converted into kinetic energy. Star formation takes place only in the most massive minihaloes, with $10$ Pop III.1 star formation sites residing in the volume that is destined to form the galaxy. This has important consequences for the role of stellar feedback, and will be further discussed in Section~4. The morphology of the galaxy can best be seen in Fig.~3, where we show a three-dimensional rendering of the central $150~\rm{kpc}$ (comoving), i.e. the same field of view as in Fig.~2. The temperature is colour-coded such that the hottest regions with $T\simeq 10^{4}~\rm{K}$ are displayed in bright red. Here, the true spacial structure of the galaxy becomes more apparent, showing that its environment is organized into prominent filaments with a high amount of substructure. In some instances, star-forming minihaloes have aligned along these filaments and will soon merge with the galaxy.

\begin{figure*}
\begin{center}
\resizebox{16cm}{16cm}
{\includegraphics{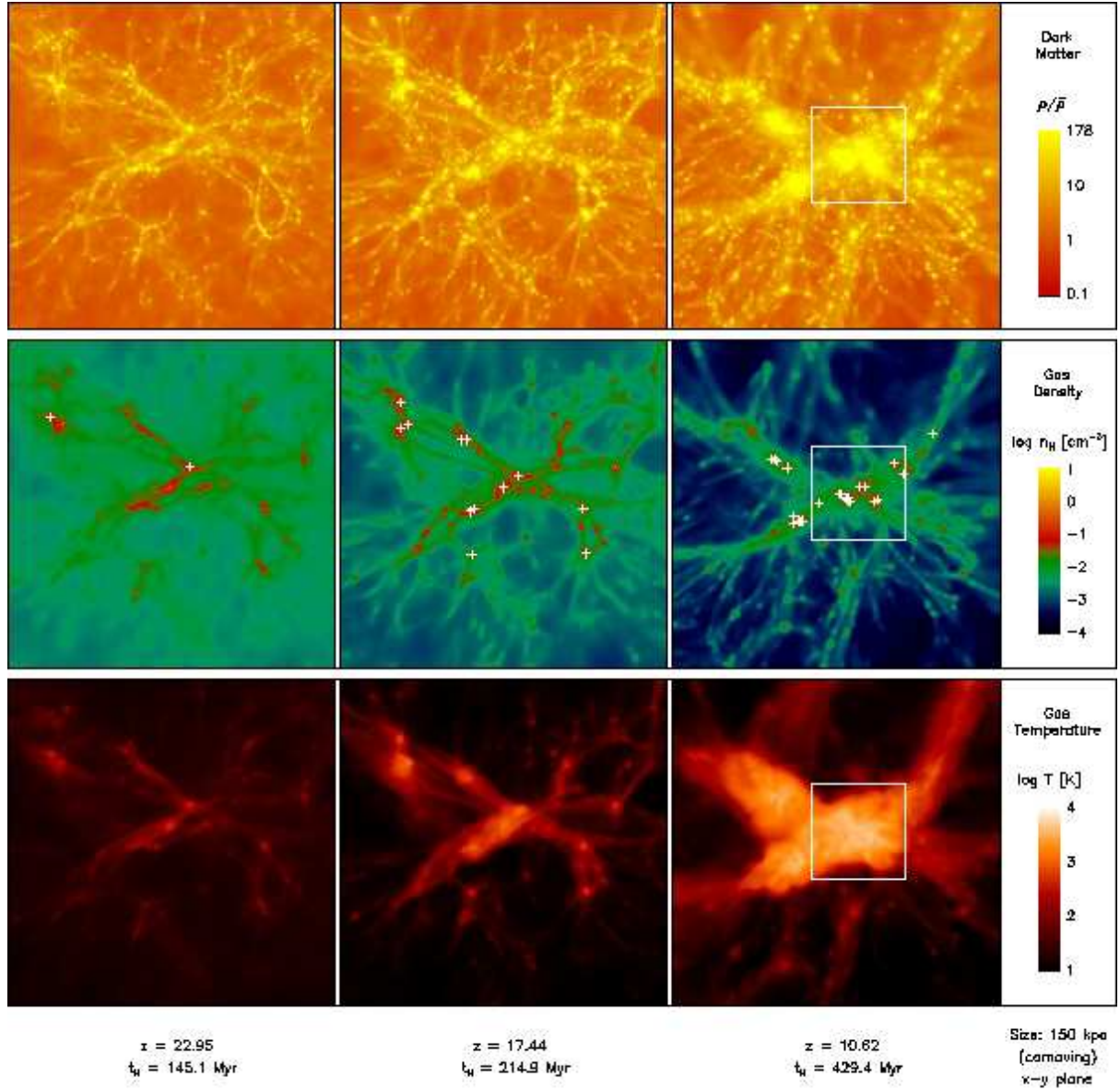}}
\caption{The DM overdensity, hydrogen number density and temperature averaged along the line of sight within the central $\simeq 150~\rm{kpc}$ (comoving) at three different output times, from $z\simeq 23$, when the first star-forming minihalo at the centre of the box collapses, to $z\sim 10$, when the first galaxy forms. White crosses denote Pop~III.1 star formation sites in minihaloes, and the insets approximately delineate the boundary of the galaxy, further enlarged in Figs~8, 9, 10 and 12. {\it Top row:} The hierarchical merging of DM haloes leads to the collapse of increasingly massive structures, with the least massive progenitors forming at the resolution limit of $\simeq 10^{4}~\rm{M}_{\odot} $ and ultimately merging into the first galaxy with $\simeq 5\times 10^{7}~\rm{M}_{\odot}$. The brightest regions mark haloes in virial equilibrium according to the commonly used criterion $\rho/\bar{\rho}>178$. Although the resulting galaxy is not yet fully virialized and is still broken up into a number of sub-components, it shares a common potential well and the infalling gas is attracted towards its centre of mass. {\it Middle row:} The gas generally follows the potential set by the DM, but pressure forces prevent collapse in haloes below $\simeq 2\times 10^{4}~\rm{M}_{\odot}$ (cosmological Jeans criterion). Moreover, star formation only occurs in haloes with virial masses above $\simeq 10^{5}~\rm{M}_{\odot}$, as densities must become high enough for molecule formation and cooling. {\it Bottom row:} The virial temperature of the first star-forming minihalo gradually increases from $\simeq 10^{3}~\rm{K}$ to $\simeq 10^{4}~\rm{K}$, at which point atomic cooling sets in.}
\end{center}
\end{figure*}

\begin{figure*}
\begin{center}
\resizebox{16cm}{16cm}{\includegraphics{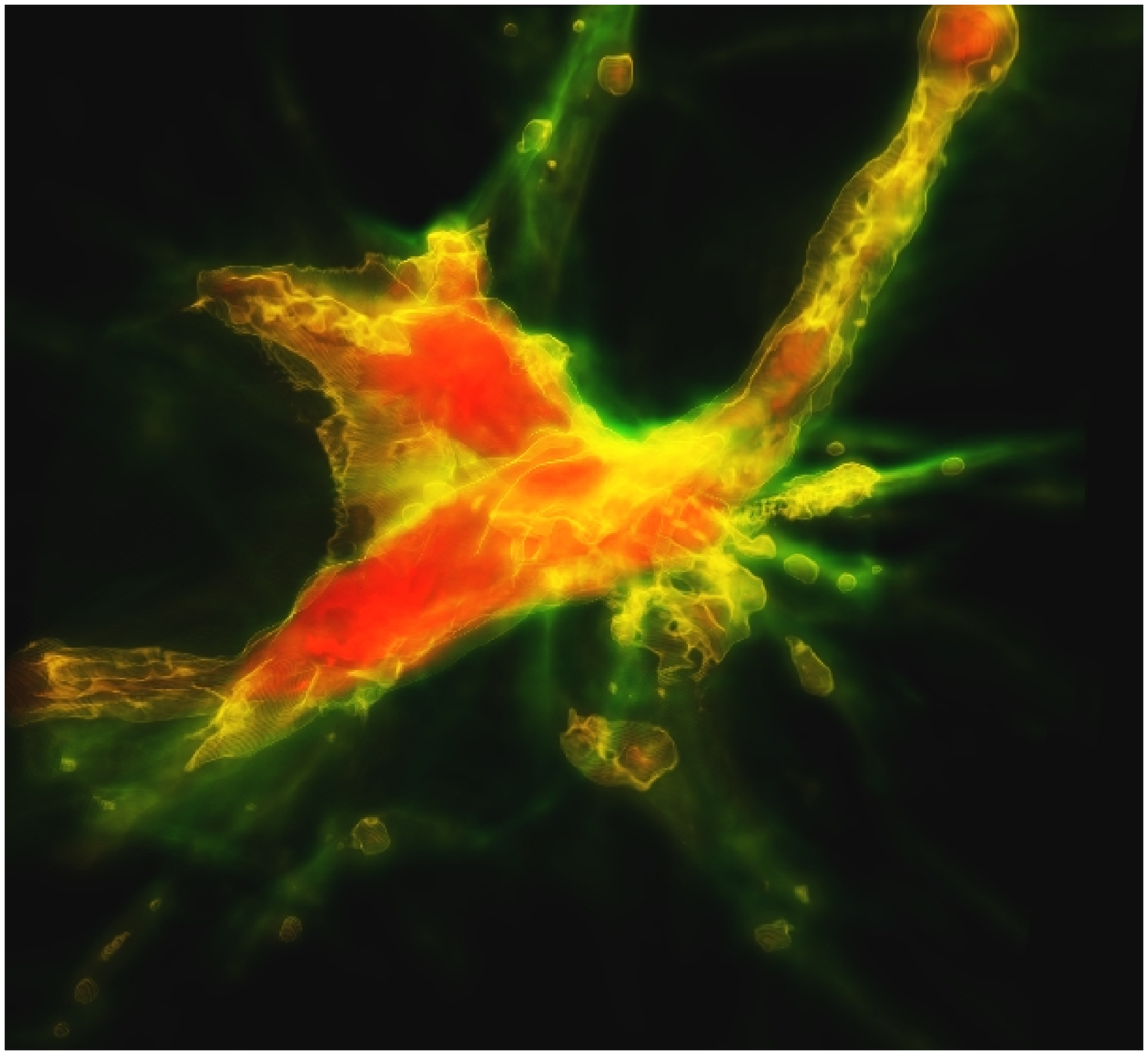}}
\caption{A three-dimensional rendering of the central $\simeq 150~\rm{kpc}$ (comoving), showing the same field of view as in Fig.~2. The temperature is colour-coded such that the hottest regions with $T\simeq 10^{4}~\rm{K}$ are displayed in bright red. Here, the true spacial structure of the galaxy becomes more clear, showing that its environment is organized into prominent filaments with a high amount of substructure. In some instances, star-forming minihaloes have aligned along these filaments and will soon merge with the galaxy.}
\end{center}
\end{figure*}

\subsection{Merger tree}
The hierarchical assembly of the galaxy can be best described by means of a merger tree that depicts the evolution of all progenitor haloes. We construct such a merger tree by tagging all DM particles that reside in the parent atomic cooling halo and track their location backwards in time. If they are part of a group at a previous timestep, they are considered to reside in a halo with mass equal to the sum of their individual masses. We repeat this process until all tagged particles are no longer part of a group or the mass falls below the resolution limit. Fig.~4 shows the resulting absolute and differential mass growth of the galaxy. Accretion is fuelled by minor as well as major mergers, with the latter showing the tendency to double the mass of the halo. In the course of $\simeq 400~\rm{Myr}$, the accretion rate increases from $\simeq 5\times 10^{-3}~\rm{M}_{\odot}~\rm{yr}^{-1}$ to $\simeq 0.5~\rm{M}_{\odot}~\rm{yr}^{-1}$, but varies significantly in between due to the highly complex nature of bottom-up structure formation.

To illustrate the substantial degree of complexity involved, Fig.~5 shows the individual paths of all progenitor haloes down to the DM resolution limit of $\simeq 10^{4}~\rm{M}_{\odot}$. The initial widening of the tree indicates that an increasing number of minihaloes collapse, while at $z\simeq 20$ merging becomes dominant and the degree of complexity decreases again. The timescale for the widening of the tree is $\simeq 150~\rm{Myr}$, while the completion of the merging process requires another $\simeq 250~\rm{Myr}$. The total number of haloes above $\simeq 10^{4}~\rm{M}_{\odot}$ that merge to form the galaxy is $\simeq 300$. A more sophisticated analysis of the merging process is presented in Fig.~6, where the individual paths and masses of all progenitor haloes are shown. Each line represents an individual halo, while the colour denotes its mass. The target halo seeding the first galaxy is indicated by the rightmost path. Sites of Pop~III.1 star formation are denoted by star symbols, and the oval denotes the formation of the atomic cooling halo. The history of this most basic building block of galaxy formation is highly complex, further complicated by the formation of $10$ Pop~III.1 stars prior to its assembly. However, this is an upper limit on previous star formation activity as we do not include radiative and SN-driven feedback, which would likely reduce the net star formation rate.

The presence of DM fluctuations on mass scales below our resolution limit might imply that Pop~III star formation takes place in haloes with viral mass well below $\simeq 2\times 10^{4}~\rm{M}_{\odot}$, but pressure forces prevent gas from settling into these shallow potential wells (cosmological Jeans criterion). Moreover, gas may not be able to collapse beyond the point of virialization in $\la 10^{5}~\rm{M}_{\odot}$ haloes, as temperatures and densities do not become high enough for efficient H$_{2}$ formation. Dynamical heating by mergers counteracts cooling and thus only a fraction of all minihaloes will be able to form stars \citep{yoshida03a}. To ascertain the importance of this effect, we determine the masses of all minihaloes experiencing star formation. As described in Section~2, the formation of a Pop~III.1 star is denoted by the creation of a sink particle once the hydrogen number density exceeds $\simeq 10^{4}~\rm{cm}^{-3}$. We implicitly assume that such a parcel of gas does not experience further subfragmentation. With this prescription, we find the following virial masses for all star-forming minihaloes shown in Fig.~6, from top to bottom: $M_{\rm{vir}}\simeq \left[5.8, 1.6, 7.5, 3.5, 4.3, 1.4, 3.2, 9.3, 1.4, 11.8\right]\times 10^{5}~\rm{M}_{\odot}$. As expected, their masses are in the range $\simeq 10^{5}$~--~$10^{6}~\rm{M}_{\odot}$, emphasizing the influence of dynamical heating on haloes below $\simeq 10^{5}~\rm{M}_{\odot}$. Interestingly, the fact that only a fraction of all minihaloes forms stars ensures a constant inflow of cold gas into existing haloes, which is crucial for the growth of the BHs at their centres.

\begin{figure}
\begin{center}
\includegraphics[width=8cm]{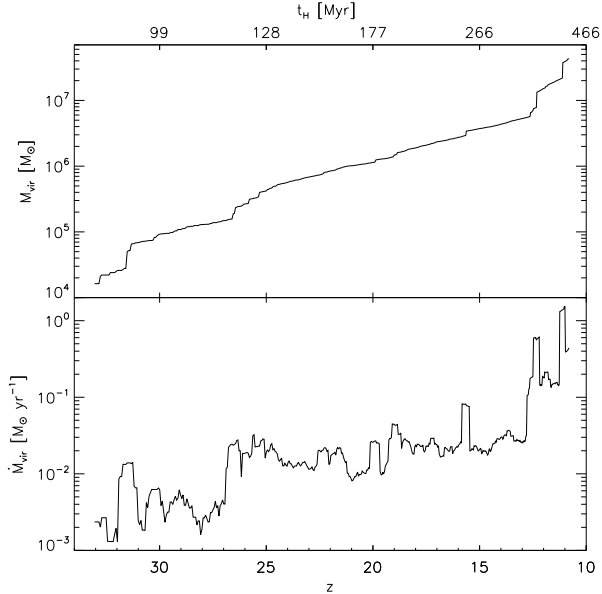}
\caption{The virial mass (top panel) and accretion rate (bottom panel) of the galaxy as a function of redshift. The growth of the underlying DM halo is fuelled by minor as well as major mergers, with the latter showing the tendency to double the mass of the target halo. At $z\simeq 10$, the atomic cooling criterion is fulfilled and a galaxy is born.}
\end{center}
\end{figure}

\begin{figure}
\begin{center}
\includegraphics[width=8cm]{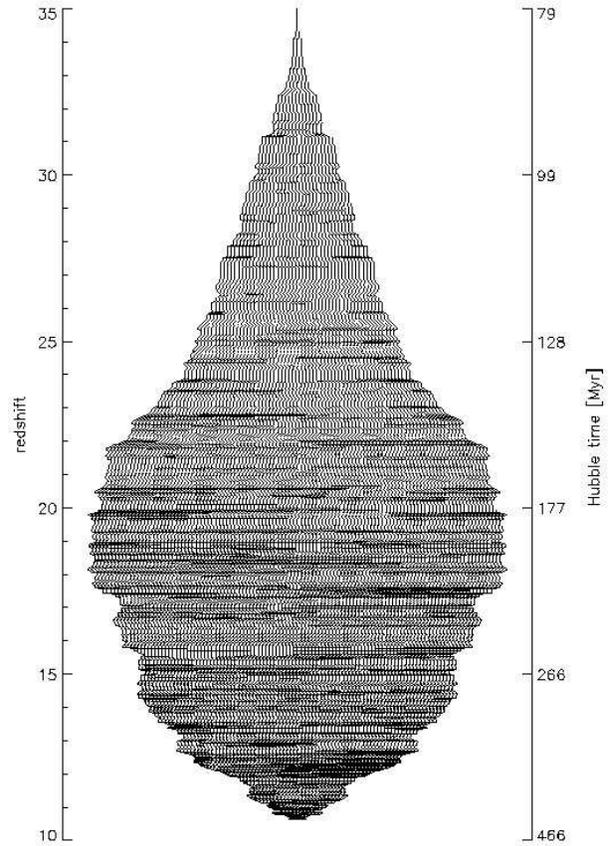}
\caption{The merger tree of the galaxy, illustrating its complexity as a function of time. New branches indicate the formation of DM haloes at the resolution limit of $\simeq 10^{4}~\rm{M}_{\odot}$. The widening of the tree increases as more and more haloes collapse, until merging dominates and the degree of complexity decreases again. The timescale for the former is $\simeq 150~\rm{Myr}$, while the completion of the merging process requires another $\simeq 250~\rm{Myr}$. A total of $\simeq 300$ haloes above $\simeq 10^{4}~\rm{M}_{\odot}$ merge to form the galaxy.}
\end{center}
\end{figure}

\begin{figure*}
\begin{center}
\resizebox{16cm}{21cm}
{\includegraphics{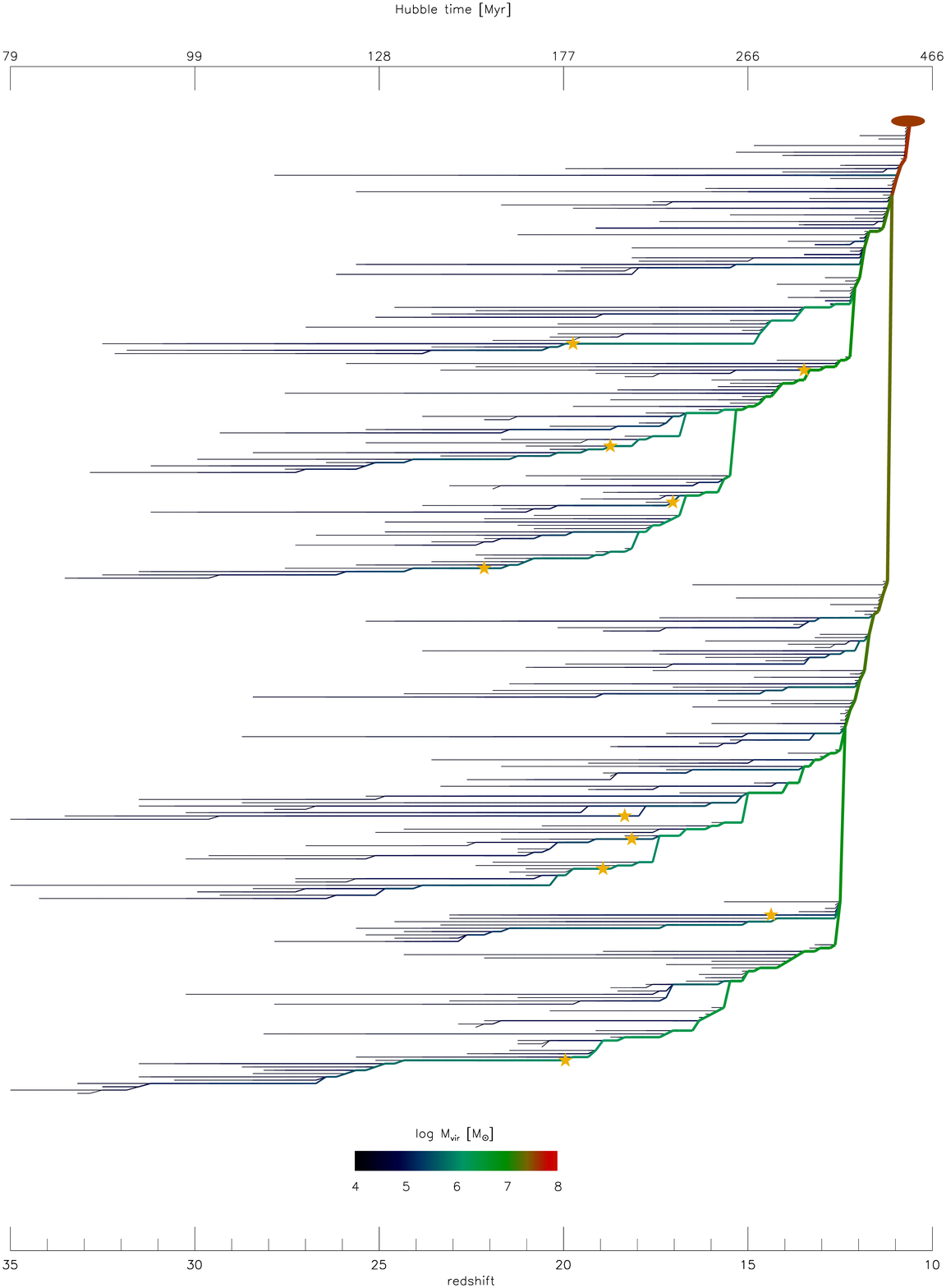}}
\caption{The full merger tree of the galaxy assembling at $z\simeq 10$. Each line represents an individual progenitor halo and is colour-coded according to its mass. The target halo seeding the galaxy is represented by the rightmost path, which ultimately attains $\simeq 5\times 10^{7}~\rm{M}_{\odot}$ and fulfils the atomic cooling criterion (denoted by the red oval). Star symbols denote the formation of Pop~III.1 stars in minihaloes, showing that in our specific realization 10 Pop~III.1 stars form prior to the assembly of the galaxy. Only a fraction of all minihaloes form stars, as dynamical heating via mergers partially offsets cooling. Depending on the detailed merger history, this ensures that star-forming minihaloes are supplied with cold gas, which is crucial for the growth of the BHs at their centres.}
\end{center}
\end{figure*}

\section{Cooling and star formation}
A crucial issue concerning the formation of the first galaxies is the chemical and thermal evolution of accreted gas, which ultimately determines the mode of star formation. We here briefly discuss radiative and SN-driven feedback exerted by the very first stars, followed by a discussion of the chemistry and cooling properties of an atomic cooling halo and the implications for second-generation star formation.

\begin{figure*}
\begin{center}
\resizebox{17cm}{9.5cm}{\includegraphics{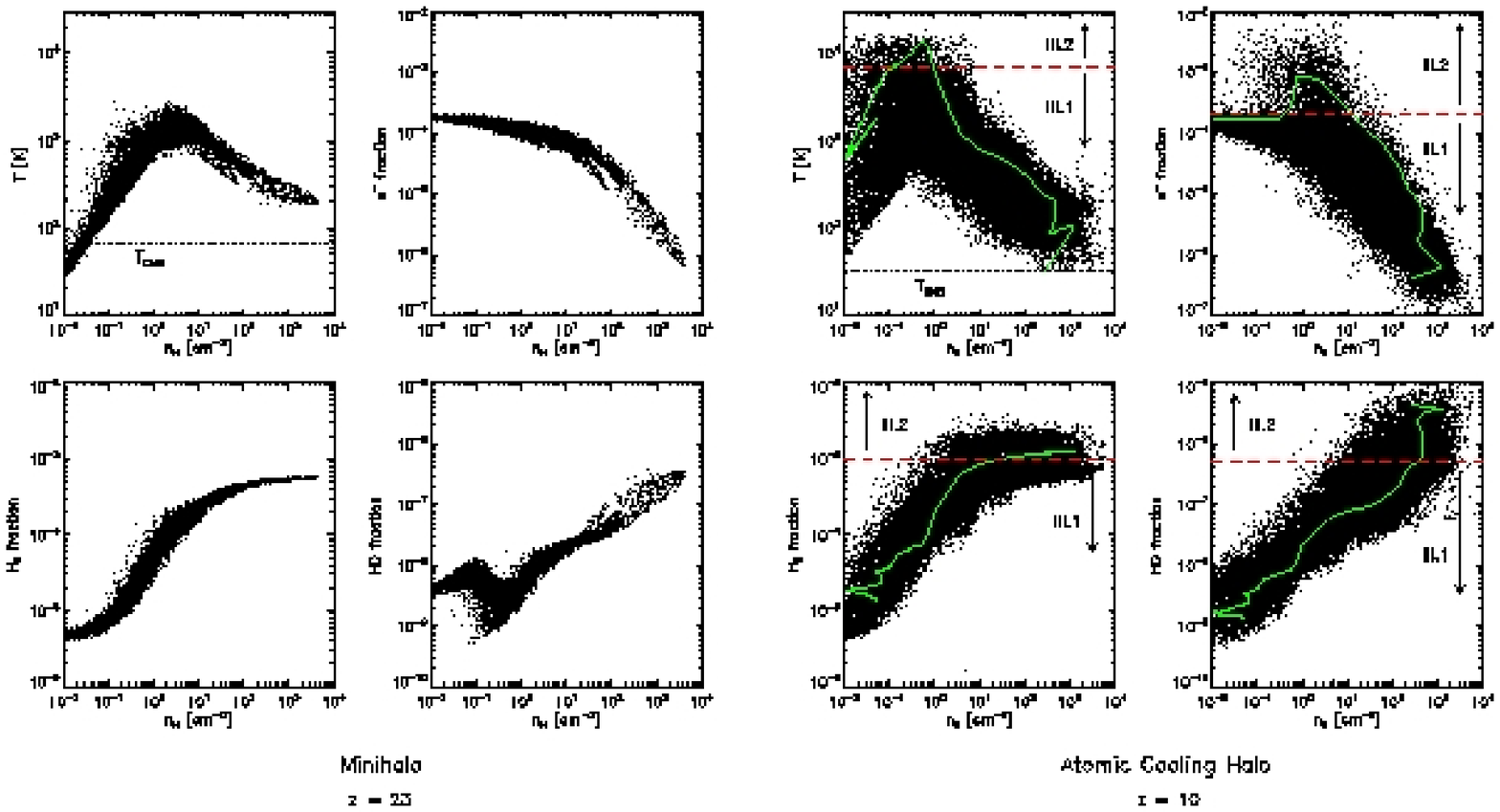}}
\caption{The phase-space distribution of gas inside the first star-forming minihalo (left-hand panel) and the atomic cooling halo (right-hand panel). We show the temperature, electron fraction, HD fraction and H$_{2}$ fraction as a function of hydrogen number density, clockwise from top left to bottom left. {\it Left-hand panel:} In the minihalo case, adiabatic collapse drives the temperature to $\ga 10^{3}~\rm{K}$ and the density to $n_{\rm{H}}\ga 1~\rm{cm}^{-3}$, where molecule formation sets in allowing the gas to cool to $\simeq 200~\rm{K}$. At this point, the central clump becomes Jeans-unstable and ultimately forms a Pop~III.1 star. {\it Right-hand panel:} In the first galaxy, a second cooling channel has emerged due to an elevated electron fraction at the virial shock, which in turn enhances molecule formation and allows the gas to cool to the temperature of the CMB. The dashed red lines and arrows approximately delineate the resulting Pop~III.1 and Pop~III.2 channels, while the solid green lines denote the path of a representative fluid element that follows the Pop~III.2 channel.}
\end{center}
\end{figure*}

\subsection{Population~III.1}
As shown in Section~3, star formation ensues in minihaloes before the larger potential wells of the first galaxies assemble. This implies that radiative and SN-driven feedback influences star formation in other minihaloes as well as second-generation star formation in the resulting atomic cooling halo. However, recent numerical simulations have shown that local radiative feedback via photoheating and LW radiation may not be as important as previously thought \citep{as07,jgb07a,whalen07}, and that a global LW background may only reduce the number of Pop~III.1 stars by $\sim 50$ per cent \citep{gb06, jgb07b}. An unknown fraction of these stars end their lives as energetic SNe and enrich the surrounding IGM to well above the critical metallicity \citep{byh03,greif07,wa07c}, while others collapse directly to BHs and do not expel any metals \citep{hw02,heger03}. Since the timescale for the recollapse of enriched gas is $\ga 100~\rm{Myr}$ \citep{greif07}, and mixing is inefficient with respect to pre-established overdensities \citep{cr08}, subsequent star formation in minihaloes prior to the assembly of the atomic cooling halo likely remains metal-free. It is much more difficult, however, to predict the character of star formation inside the first galaxies. In the following, we will first examine the consequences of pristine gas collapsing in the atomic cooling halo, and subsequently briefly address the corresponding case of pre-enriched gas.

\subsection{Population~III.2}
The possible existence of a distinct population of metal-free stars in regions of previous ionization has attracted increasing attention (\citealt{mbh03,gb06,jb06,yoshida06a,tm08,yoshida07}; but see \citealt{ripamonti07,mb08}). According to theory, an elevated electron fraction catalyses the formation of H$_{2}$ and HD well above the level found in minihaloes and enables the gas to cool to the temperature of the CMB. This reduces the Bonnor-Ebert mass by almost an order of magnitude and likely leads to the formation of Pop~III.2 stars with $\ga 10~\rm{M}_{\odot}$ \citep{jb06}. Numerical simulations of star formation in relic H~{\sc ii} regions have largely confirmed this picture \citep{jgb07a,yoh07}, while its relevance during the virialization of the first galaxies has not yet been established \citep[but see][]{gb06,jgb07b}.

The chemistry of gas contracting in an atomic cooling halo is fundamentally different from that in minihaloes. The latter maintain a primordial electron fraction of $\simeq 3\times 10^{-4}$ and form a limited amount of molecules, while the virial temperature in an atomic cooling halo exceeds $\simeq 10^{4}~\rm{K}$ and the elevated electron fraction facilitates the formation of high H$_{2}$ and HD abundances. This allows the gas to cool to the temperature of the CMB instead of the canonical $\simeq 200~\rm{K}$ found in minihaloes. In the left panel of Fig.~7, we show the properties of the gas in the first star-forming minihalo at the centre of the computational box. The primordial electron fraction remains constant until densities become high enough for electron recombination. After adiabatic heating to $\ga 10^{3}~\rm{K}$, molecule formation sets in and the gas cools to $\simeq 200~\rm{K}$, at which point the central clump becomes Jeans-unstable and inevitably forms a Pop~III.1 star. In contrast, the right panel of Fig.~7 shows the density and temperature of the gas inside the atomic cooling halo. The conventional H$_{2}$ cooling channel is still visible, but a second path from low to high density has emerged, enabled by an elevated electron fraction at the virial shock, which in turn enhances the formation of H$_{2}$ and HD and allows the gas to cool to the temperature of the CMB. The dashed red lines and black arrows in Fig.~7 approximately delineate both channels, showing that the electron fraction in the Pop~III.2 case is elevated by an order of magnitude to $\sim 10^{-3}$, and the H$_{2}$ fraction rises to $\simeq 2\times 10^{-3}$. As already estimated in \citet{jgb07b}, the HD fraction grows to above $\simeq 10^{-6}$.

To more clearly illustrate this point, we plot the path of a representative fluid element evolving along the Pop~III.2 channel (solid green lines in Fig.~7). Such gas indeed cools to the CMB floor, potentially enabling the formation of Pop~III.2 stars. The mass fraction entering this channel is relatively low at the time considered here since the atomic cooling threshold has just been surpassed, but should quickly rise as freshly accreted material is shock-heated to $\simeq 10^{4}~\rm{K}$. We cannot study any possible fragmentation, since the gas rapidly falls to within the Bondi radius of the central BH, and is thus accreted by the sink particle. It will be very interesting to investigate the fragmentation of the Pop~III.2 mode in future, higher-resolution simulations, in particular testing the predicted mass scale of $\ga 10~\rm{M}_{\odot}$ \citep[see e.g.][]{cgk08}.

What are the implications of this result? Parcels of gas that are accreted on to the galaxy through the virial shock can cool to the temperature of the CMB and possibly become gravitationally unstable, resulting in the formation of Pop~III.2 stars. Including radiative feedback from previous star formation would only strengthen this conclusion, as the degree of ionization would be increased even further \citep{wa07c}. As long as the gas collapsing into the first galaxies remains pristine, primordial star formation will therefore likely be dominated by intermediate-mass (Pop~III.2) stars \citep{gb06}. The crucial question, however, is: Can the gas inside the first galaxies remain metal-free?

\subsection{Population~II}
In the previous sections, we have found that of order $10$ Pop~III.1 stars form prior to the assembly of the atomic cooling halo. In this case it appears unlikely that all of them will collapse into BHs without any metal-enrichment \citep{jgb07b}. Even a single SN from a massive Pop~III star would already suffice to reach levels above the critical metallicity, at least on average \citep{byh03,greif07,wa07c}. More generally, at some stage in cosmic history, there must have been a transition from primordial, high-mass star formation to the 'normal' mode that dominates today. The discovery of extremely metal-poor stars in the Galactic halo with masses below one solar mass \citep{christlieb02,frebel05,bc05} indicates that this transition occurs at abundances considerably smaller than the solar value. At the extreme end, these stars have iron abundances less than $10^{-5}$ times the solar value, but show significant carbon and oxygen enhancements, which could be due to unusual abundance patterns produced by enrichment from BH-forming Pop~III SNe \citep{un03}, or due to mass transfer from a close binary companion, whose frequency is predicted to increase with decreasing metallicity \citep{lucatello05}.

Identifying the critical metallicity at which this transition occurs is subject to ongoing research. One approach is to argue that low mass star formation becomes possible only when atomic fine-structure line cooling from carbon and oxygen becomes effective \citep{bromm01,bl03a,ss06a,fjb07}, setting a value for $\rm{Z_{crit}}$ at $\simeq 10^{-3.5}~\rm{Z}_{\odot}$. Another possibility, first proposed by \citet{omukai05}, is that low mass star formation is a result of dust-induced fragmentation occurring at high densities, $n_{\rm{H}}\simeq 10^{13}~\rm{cm}^{-3}$, and thus at a very late stage in the protostellar collapse. In this model, $10^{-6}\la \rm{Z_{crit}}\la 10^{-5}~\rm{Z}_{\odot}$, where much of the uncertainty in the predicted value results from uncertainties in the dust composition and the degree of gas-phase depletion \citep{schneider02,schneider06}. Recent numerical simulations by \citet{to06} as well as \citet{cgk08} provide support for this picture. However, the existing data of metal-poor Galactic halo stars seems to be well accommodated by the C- and O-based fine-structure model \citep{fjb07}.

In the present simulation, we do not follow the metallicity evolution of the infalling gas. Thus, we can only speculate about the properties of the resulting stellar population. It appears reasonable to assume that some of the accreting material is still pristine and free of metals, triggering the formation of lower-mass metal-free Pop~III.2 stars. Gas that flows in at even later times may already have experienced metal enrichment from previous Pop~III SNe in nearby minihaloes. Because of the high level of turbulence within the virial radius at that time (see Section~5), the incoming new material is likely to efficiently mix with the pre-existing zero-metallicity gas and the era of Pop~III star formation could come to an end. This transition possibly occurs at the same time as the onset of significant degrees of turbulence in the atomic cooling halo. We therefore speculate that some of the extremely metal-deficient stars in the halo of the Milky Way may have formed as early as redshift $z\simeq 10$ \citep{cgk08}.

\section{Turbulence}
The development of turbulence in gas flowing into the central potential well of the halo strongly influences its fragmentation behaviour and consequently its ability to form stars. Detailed studies of the interstellar medium in the Milky Way, for instance, tell us that turbulence determines when and where star formation occurs and that it is the intricate interplay between gravity on the one hand, and turbulence, thermal pressure and magnetic fields on the other that sets the properties of young stars and star clusters \citep{larson03,mk04,ballesteros-paredes07}. In the context of our work, we investigate the velocity field and energy distribution that build up during the assembly of the galaxy. As opposed to cooling flows in low-mass haloes, the accretion flow on to the deep central potential well of the atomic cooling halo considered here becomes highly turbulent within the virial radius.

\subsection{The development of turbulence: hot versus cold accretion}
One of the most important consequences of atomic cooling is the softening of the equation of state below the virial radius, allowing a fraction of the potential energy to be converted into kinetic energy \citep{wa07b}. This implies that perturbations in the gravitational potential can generate turbulent motions on galactic scales, which are then transported to the centre of the galaxy (e.g. Fig.~8). In this context it is important to investigate the accretion of gas on to the galaxy in more detail.

In principle, there are two distinct modes of accretion. Gas accreted directly from the IGM is heated to the virial temperature and comprises the sole channel of inflow until cooling in filaments becomes important. This mode is termed hot accretion, and dominates in low-mass haloes at high redshift. In the atomic cooling halo, the formation of the virial shock and the concomitant heating are visible in Fig.~9, where we show the hydrogen number density and temperature of the central $\simeq 40~\rm{kpc}$ (comoving) around the BH at the centre of the galaxy. This case also reveals a second mode, termed cold accretion. It becomes important as soon as filaments are massive enough to enable molecule reformation, which allows the gas to cool and flow into the central regions of the nascent galaxy with high velocities. In Fig.~9, the cold gas accreted along filaments from the left- and right-hand is clearly distinguishable from the hot gas at the virial shock. These streams are also visible in Fig.~10, where we compare the radial with the tangential velocity component, and in Fig.~8, where we show the Mach number of infalling gas. Evidently, inflow velocities can be as high as $20~\rm{km}~\rm{s}^{-1}$, with Mach numbers of the order of $10$.

In Fig.~11, we compare the energy distribution and mass fraction of cold ($<500~\rm{K}$) versus hot ($>500~\rm{K}$) gas in radial shells for the first star-forming minihalo just before the formation of the sink particle, and the atomic cooling halo assembling at $z\simeq 10$. The blue, green and red lines denote the azimuthally averaged ratio of radial, tangential and thermal to potential energy, respectively. The black lines show the sum of all three components, and the dotted lines indicate the ratio required for perfect virialization. In the minihalo case, the total energy is dominated by thermal energy, although its share decreases towards the centre where cooling via molecular hydrogen becomes important. The radial kinetic energy dominates over the tangential component down to $r_{\rm{tan}}\simeq 5~\rm{pc}$, where the mass fraction of cold gas rapidly rises and the cloud becomes rotationally supported. Efficient cooling implies that the total energy drops below that required for perfect virialization.

In the atomic cooling halo, the total energy at $r_{\rm{vir}}\simeq 1~\rm{kpc}$ is dominated by bulk radial motions. The distinction between hot and cold gas in the right panels of Fig.~11 shows that a large fraction of the kinetic energy injected into the galaxy comes from cold gas accreted along filaments, even though its mass fraction is initially small. The energy in tangential motions begins to dominate at $r_{\rm{tan}}\simeq 200~\rm{pc}$, showing that the radial energy of the cold gas flowing in along filaments is converted into turbulent motions. This is fundamentally different from the collapse of gas in minihaloes, where the radial energy is converted into a directed rotation along a single axis. The distinct features at $r\simeq 350~\rm{pc}$ are caused by a subhalo that has not yet merged with the central clump (see also Figs~2 and 3). Once again, the total energy budget falls below that required for perfect virialization, as atomic hydrogen as well as molecular cooling are able to radiate away a significant fraction of the potential energy released. We conclude that the high energy input by cold accretion is ideally suited to drive turbulence at the centre of the galaxy, where bulk radial inflows are converted into turbulent motions on small scales.

\begin{figure}
\begin{center}
\includegraphics[width=8cm]{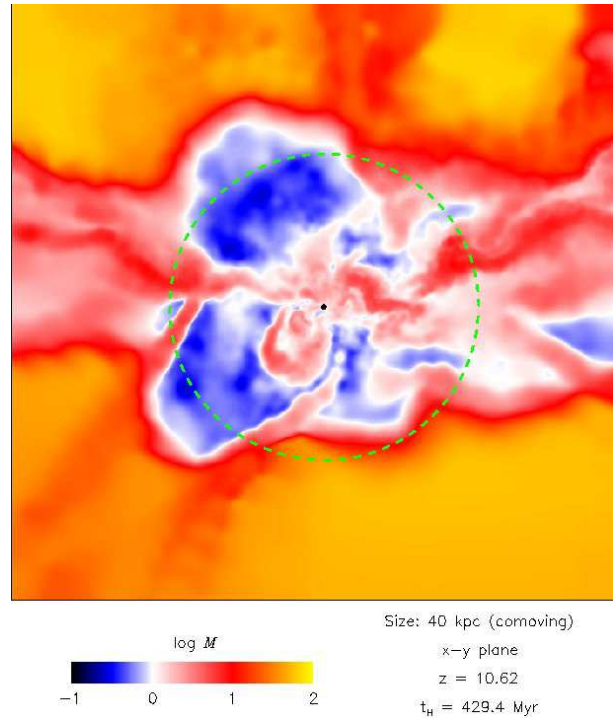}
\caption{The central $\simeq 40~\rm{kpc}$ (comoving) of the computational box, roughly delineated by the insets in Fig.~2. Shown is the Mach number in a slice centred on the BH at the centre of the galaxy, indicated by the filled black circle. The dashed line denotes the virial radius at a distance of $\simeq 1~\rm{kpc}$.  The Mach number approaches unity at the virial shock, where gas accreted from the IGM is heated to the virial temperature over a comparatively small distance. Inflows of cold gas along filaments are supersonic by a factor of $\simeq 10$ and generate a high amount of turbulence at the centre of the galaxy, where typical Mach numbers are between $1$ and $5$.}
\end{center}
\end{figure}

\begin{figure*}
\begin{center}
\resizebox{16cm}{9.5cm}{\includegraphics{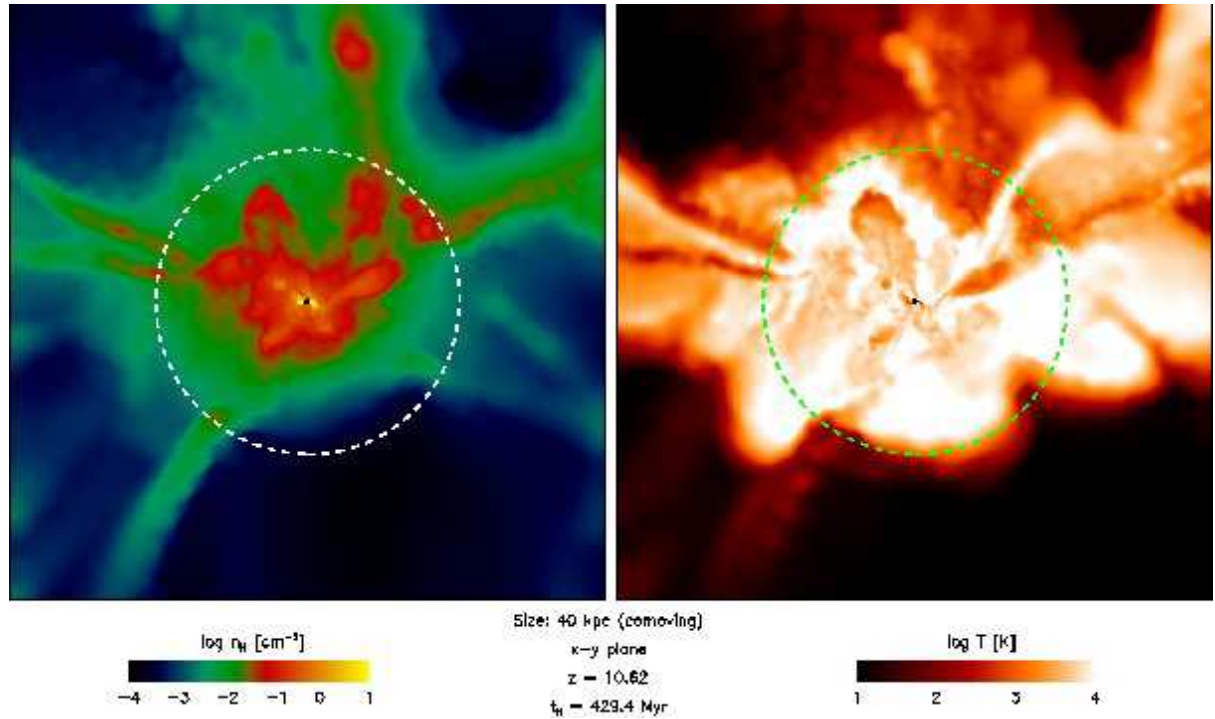}}
\caption{The central $\simeq 40~\rm{kpc}$ (comoving) of the computational box, roughly delineated by the insets in Fig.~2. Shown is the hydrogen number density (left-hand panel) and temperature (right-hand panel) in a slice centred on the BH at the centre of the galaxy, indicated by the filled black circle. The dashed lines denote the virial radius at a distance of $\simeq 1~\rm{kpc}$. Hot accretion dominates where gas is accreted directly from the IGM and shock-heated to $\simeq 10^{4}~\rm{K}$. In contrast, cold accretion becomes important as soon as gas cools in filaments and flows towards the centre of the galaxy, such as the streams coming from the left- and right-hand side. They drive a prodigious amount of turbulence and create transitory density perturbations that could in principle become Jeans-unstable. In contrast to minihaloes, the initial conditions for second-generation star formation are highly complex, with turbulent velocity fields setting the fragmentation properties of the gas.}
\end{center}
\end{figure*}

\begin{figure*}
\begin{center}
\resizebox{16cm}{9.5cm}{\includegraphics{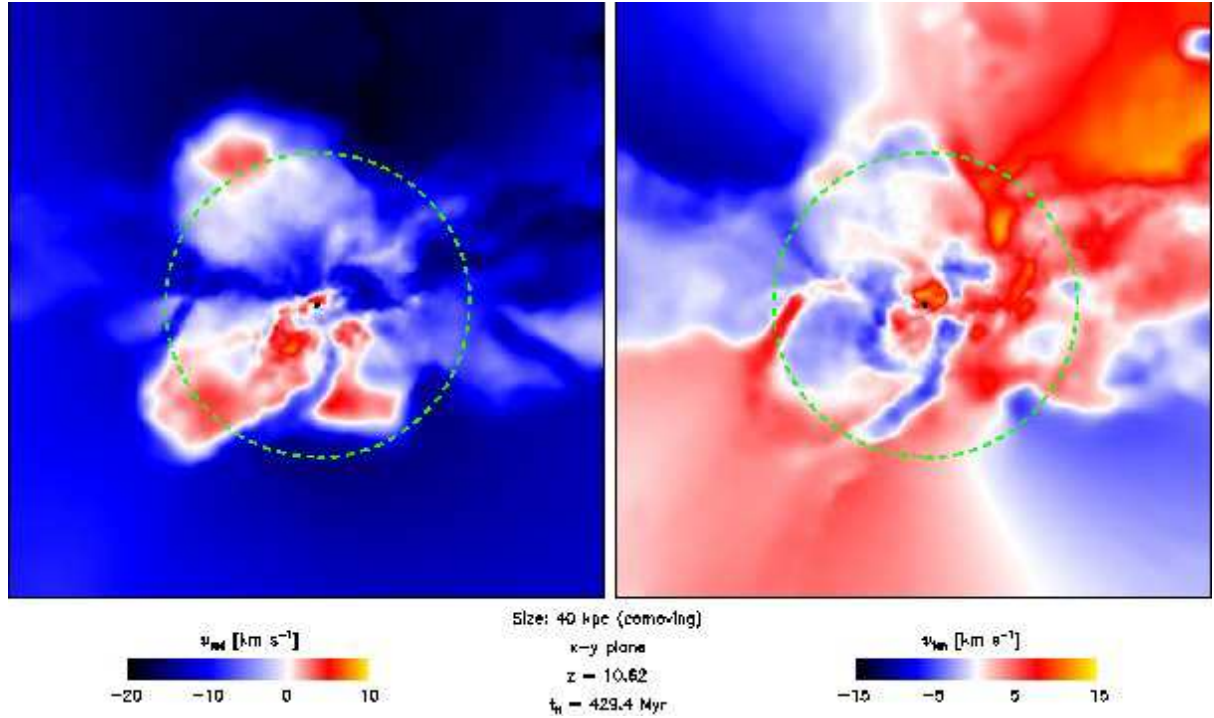}}
\caption{The central $\simeq 40~\rm{kpc}$ (comoving) of the computational box, roughly delineated by the insets in Fig.~2. Shown is the radial (left-hand panel) and tangential velocity in the x-y plane (right-hand panel) in a slice centred on the BH at the centre of the galaxy, indicated by the filled black circle. The dashed lines denote the virial radius at a distance of $\simeq 1~\rm{kpc}$. Streams of cold gas from filaments, such as those coming from the left- and right-hand side, are clearly visible and can have velocities of up to $20~\rm{km}~\rm{s}^{-1}$. Some even penetrate the centre and create regions of positive radial velocities. Angular velocities are particularly high towards the centre of the galaxy, where bulk radial inflows are converted into turbulent motions on small scales. The presence of flows in both directions implies that these are unorganized instead of coherently rotating, such as is the case in minihaloes (see also Fig.~11).}
\end{center}
\end{figure*}

\begin{figure*}
\begin{center}
\resizebox{16cm}{10cm}{\includegraphics{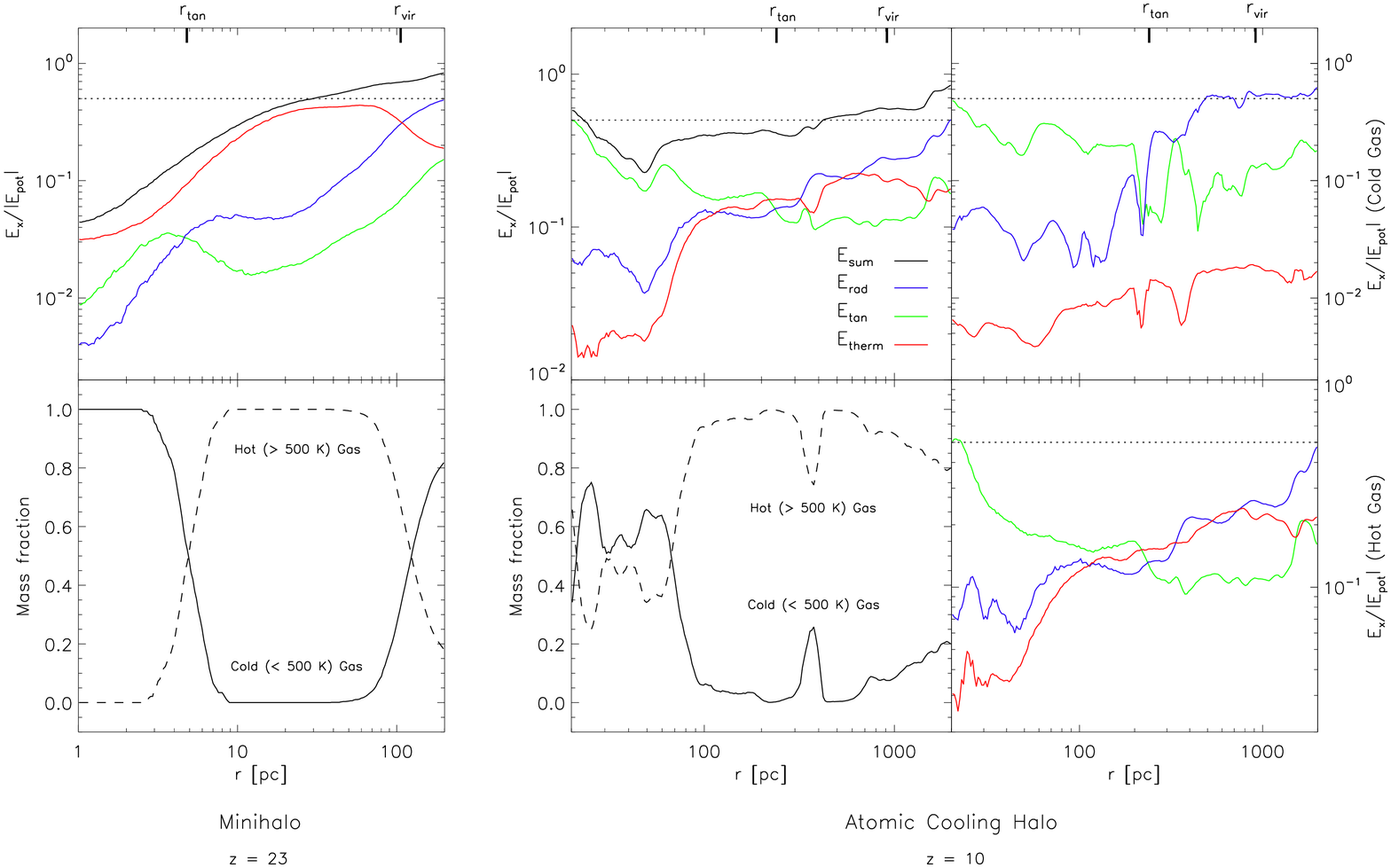}}
\caption{The energy distribution and mass fraction of cold ($<500~\rm{K}$) versus hot ($>500~\rm{K}$) gas in radial shells for the first star-forming minihalo just before the formation of the sink particle, and the atomic cooling halo assembling at $z\simeq 10$. The azimuthally averaged ratio of radial, tangential and thermal to potential energy are shown as blue, green and red lines, respectively. The black lines show the sum of all three components, and the dotted lines indicate the ratio required for perfect virialization. In the minihalo case, the total energy is dominated by thermal energy, although its share decreases towards the centre where cooling via molecular hydrogen becomes important. The radial kinetic energy dominates over the tangential component down to $r_{\rm{tan}}\simeq 5~\rm{pc}$, where the mass fraction of cold gas rapidly rises and the cloud becomes rotationally supported. In the atomic cooling halo, the total energy at $r_{\rm{vir}}\simeq 1~\rm{kpc}$ is dominated by bulk radial motions. The distinction between hot and cold gas in the right panels shows that most of the kinetic energy injected into the galaxy comes from the cold gas accreted along filaments, even though its mass fraction is initially small. The tangential component begins to dominate at $r_{\rm{tan}}\simeq 200~\rm{pc}$, where the radial flow of cold gas along filaments is converted into turbulent motions. The distinct features at $r\simeq 350~\rm{pc}$ are caused by a subhalo that has not yet merged with the central clump (see also Figs~2 and 3).}
\end{center}
\end{figure*}

\subsection{Shocks and fragmentation properties}
Shock fronts can arise where supersonic flows experience sudden deceleration and, if unorganized, indicate the presence of supersonic turbulence. As discussed above, cold accretion is a viable agent for driving turbulence, due to the prodigious amount of momentum and kinetic energy it brings to the centre of the galaxy. In Fig.~12, we show the divergence and vorticity of the velocity field during the virialization of the galaxy. A comparison with Fig.~8 implies that there are indeed regions of supersonic flow that experience rapid deceleration and form shocks. In our case, two physically distinct mechanisms are responsible for creating these shocks. The virial shock forms where the ratio of infall velocity to local sound speed approaches unity, and is clearly visible in the left-hand panel of Fig.~12. The velocity divergence is negative since the gas rapidly decelerates, while the vorticity is almost negligible. In contrast, the unorganized multitude of shocks that form near the centre of the galaxy are mostly caused by accretion of cold, high-velocity gas from filaments. These are more pronounced than the virial shock and have a significantly higher angular component. They create transitory density perturbations that could in principle become Jeans-unstable and trigger the gravitational collapse of individual clumps.

How does the turbulence generated in the infalling material influence its fragmentation behaviour and control subsequent star formation? From detailed observational and theoretical studies of star formation in our Milky Way we know that turbulence plays a pivotal role in the formation of stars and star clusters. It is usually strong enough to counterbalance gravity on global scales. By the same token, however, it will usually provoke collapse locally. Turbulence establishes a complex network of interacting shocks, where regions of high density build up at the stagnation points of convergent flows. To result in the formation of stars, local collapse must progress to high enough densities on time scales shorter than the typical interval between two successive shock passages. Only then can the collapsing core decouple from the ambient flow pattern and build up a star. The accretion flow on to these objects and consequently the final stellar mass strongly depends on the properties of the surrounding turbulent flow.

In concert with the thermodynamic properties of the gas, leading to the cooling of high-density material to the CMB limit (see Section~4), length scale and strength of the turbulence are the most important parameters governing its fragmentation behaviour and consequently the properties of star formation, such as its timescale and overall efficiency \citep{khm00,vbk03,km05a}. In the atomic cooling halo discussed here, this will eventually lead to the transition to Pop~II star formation. However, a quantitative understanding of the fragmentation behaviour of the turbulent gas would require dedicated high-resolution simulations, which is beyond the scope of this work.

\begin{figure*}
\begin{center}
\resizebox{16cm}{9.5cm}{\includegraphics{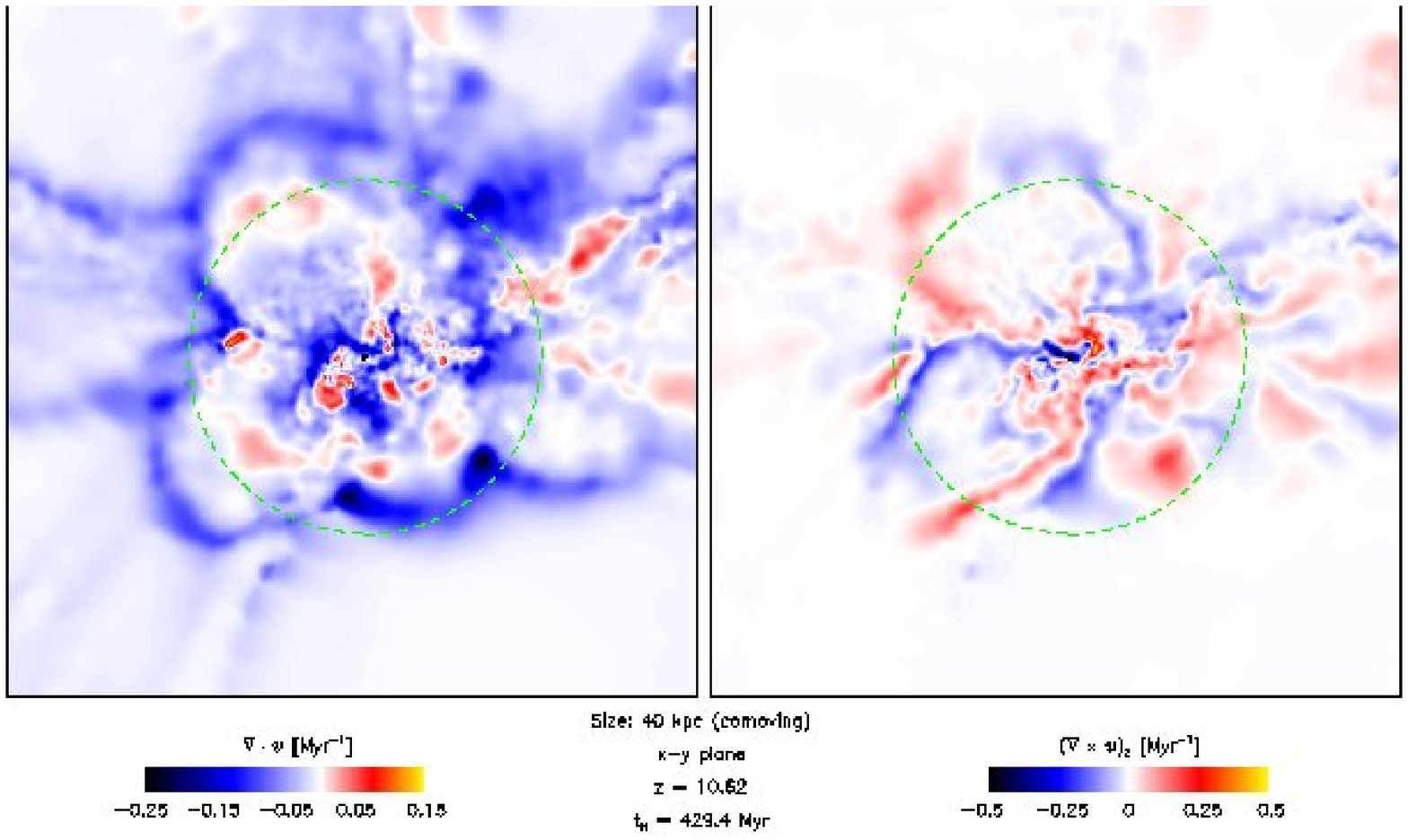}}
\caption{The central $\simeq 40~\rm{kpc}$ (comoving) of the computational box, roughly delineated by the insets in Fig.~2. We show the divergence (left-hand panel) and z-component of the vorticity (right-hand panel) in a slice centred on the BH at the centre of the galaxy, indicated by the filled black circle. The dashed lines denote the virial radius at a distance of $\simeq 1~\rm{kpc}$. The most pronounced feature in the left-hand panel is the virial shock, where the ratio of infall speed to local sound speed approaches unity and the gas decelerates over a comparatively small distance. In contrast, the vorticity at the virial shock is almost negligible. The high velocity gradients at the centre of the galaxy indicate the formation of a multitude of shocks where the bulk radial flows of filaments are converted into turbulent motions on small scales.}
\end{center}
\end{figure*}

\section{Massive black hole growth}
Galaxy formation in general involves the co-evolution of a central black hole and the surrounding stellar system, the one influencing the other. Two crucial unsolved problems are: What were the seeds for BH growth, and how important was this co-evolution at very high redshifts? We here begin to address these questions. Different scenarios have been suggested to account for the seeds of BH growth \citep{rees84}: the direct collapse of gas in atomic cooling haloes in the presence of a strong photodissociating background, or stellar remnants of massive, metal-free stars. In the following, we discuss the latter possibility and investigate the growth of a MBH forming at the centre of the galaxy.

\subsection{Accretion rate}
Even though studies of stellar evolution have shown that primordial stars may explode as energetic SNe, we here assume that all Pop~III.1 stars collapse directly to BHs \citep{hw02,heger03}. Their initial mass is dictated by the resolution limit to $M_{\rm{BH}}\simeq 2\times 10^{3}~\rm{M}_{\odot}$, with accretion on to the BH governed by the criteria discussed in Section~2. Recent investigations have shown that photoheating by the progenitor star can delay efficient accretion by reducing the central density to $\la 1~\rm{cm}^{-3}$ \citep{jb07}. However, the suppression of accretion also depends on the detailed merger history of the host halo. For example, a major merger occuring just after the formation of the BH could transport enough cold gas to its centre to enable accretion at the Eddington rate. As shown in Fig.~6, some minihaloes merge shortly after forming a BH, in some cases after only a few million years. On the other hand, if mergers are absent or incoming haloes are not sufficiently massive, accretion could be suppressed for $\ga 100~\rm{Myr}$ \citep{jb07}. In our approach to derive an upper limit on the accretion rate, we neglect the effects of photoheating by the progenitor star  such that accretion is governed solely by the supply of cold gas brought to the centre of the halo.

A more precise modelling would also require a prescription for radiation emitted by the BH-powered miniquasar and its feedback on the surrounding disc \citep{vr06}. In a simpler approach, one can assume a given radiative efficiency $\epsilon$, which denotes the ratio of BH luminosity to accreted mass energy, and assume that accretion is spherically symmetric. This leads to the Eddington accretion rate:
\begin{equation}
\dot{M}_{\rm{Edd}}=\frac{1}{\epsilon}\frac{M_{\rm{BH}}}{t_{\rm{Salp}}}\mbox{\ ,}
\end{equation}
where $t_{\rm{Salp}}$ is the Salpeter time, defined by:
\begin{equation}
t_{\rm{Salp}}=\frac{c\sigma_{\rm{T}}}{4\pi Gm_{\rm{H}}}\simeq 450~\rm{Myr}\mbox{\ .}
\end{equation}
The mass of the BH as a function of time is thus given by:
\begin{equation}
M_{\rm{BH}}(t)=M_{\rm{BH}}(t_{0})\;\rm{exp}\left(\frac{1-\epsilon}{\epsilon}\;\frac{t-t_{0}}{t_{\rm{Salp}}}\right)\mbox{\ .}
\end{equation}
In Fig.~13, we compare the mass growth of the most massive BH with the Eddington-limited model, using a fiducial value of $\epsilon=1/10$. We find that the accretion rate remains roughly constant at $\simeq 5\times 10^{-3}~\rm{M}_{\odot}~\rm{yr}^{-1}$, such that the BH grows from $\simeq 2\times 10^{3}$ to $\simeq 10^{6}~\rm{M}_{\odot}$ in the course of $\simeq 300~\rm{Myr}$. Due to our neglect of radiative feedback, the BH accretes well above the Eddington rate throughout most of its lifetime. At later times the accretion rate stagnates, most likely caused by the high kinetic energy input at the centre via cold accretion (see Fig.~11). Consequently, the fraction of unbound gas near the BH increases and its mass growth is slowed. We conclude that Eddington accretion is possible under the most favourable circumstances, for example, where a recent merger brings an ample supply of cold gas to the centre of the halo, but generally radiative feedback by the progenitor star and the disc around the BH will lead to sub-Eddington accretion rates \citep{jb07,pdc07}.

\begin{figure}
\begin{center}
\includegraphics[width=8cm]{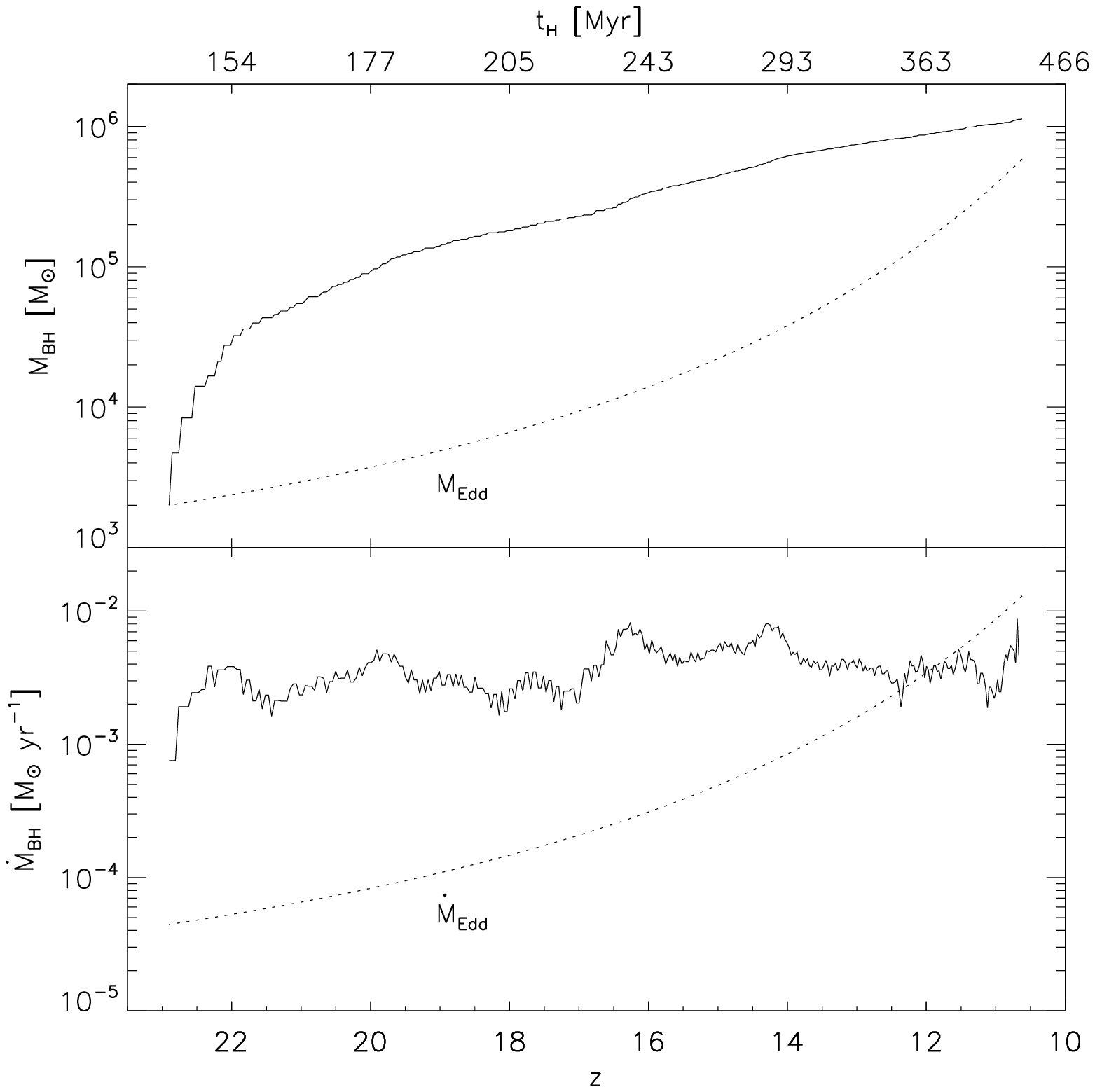}
\caption{The mass (top panel) and accretion rate (bottom panel) of the central BH as a function of redshift, shown for the simulation (solid lines) and Eddington-limited accretion (dotted lines). The accretion rate remains roughly constant at $\simeq 5\times 10^{-3}~\rm{M}_{\odot}~\rm{yr}^{-1}$, such that the BH grows from $\simeq 2\times 10^{3}$ to $\simeq 10^{6}~\rm{M}_{\odot}$ in the course of $\simeq 300~\rm{Myr}$. This is a strict upper limit as radiation effects are not taken into account. Accretion is initially super-Eddington due to the high amount of cold gas brought to the centre of the galaxy, but stagnates once turbulence becomes important and the fraction of unbound gas increases.}
\end{center}
\end{figure}

\subsection{Accretion luminosity}
The radiation generated by the BH-powered miniquasar can have numerous effects on the formation of the first galaxies. In addition to its negative feedback on star formation via photoheating, the emitted radiation can contribute to the LW background, as well as to the reionization of the Universe \citep{madau04,ro04,km05b}. In the following, we derive an upper limit on the photodissociating flux and the number of ionizing photons emitted by a stellar remnant BH accreting at the Eddington limit.

\subsubsection{Photodissociating flux}
To determine the flux of LW photons, we first model the temperature profile of the surrounding accretion disc:
\begin{equation}
T(r)=\left(\frac{3}{8\pi}\frac{GM_{\rm{BH}}\dot{M}_{\rm{BH}}}{\sigma_{\rm{SB}} r^{3}}\right)^{\frac{1}{4}}\mbox{\ ,}
\end{equation}
where $r$ is the distance from the BH, $M_{\rm{BH}}$ its mass, $\dot{M}_{\rm{BH}}$ the accretion rate and $\sigma_{\rm{SB}}$ the Stefan-Boltzmann constant \citep{pringle81}. For simplicity, we have taken the disc to be a thin disc, such that each annulus radiates as a blackbody of temperature given by the above equation. The inner-most radius of the disc is given by:
\begin{equation}
r_{\rm{inner}}\sim 2~\rm{km}\left(\frac{M_{\rm{BH}}}{\rm{M}_{\odot}}\right)\mbox{\ , }
\end{equation}
corresponding to a high value for the BH spin parameter $a\ga 0.9$ \citep{makishima00,vierdayanti07}, which we expect considering the large angular momentum of the accreted gas (see Fig.~11). We integrate the flux over the surface of the disc from $r_{\rm{inner}}$ to $r_{\rm{outer}}=10^{4} r_{\rm{inner}}$, where the contributions to both the photodissociating and ionizing fluxes are negligible. To determine the flux emitted in the LW bands, we evaluate the total emitted flux at $12.87~\rm{eV}$. In the upper panel of Fig.~14, we show the LW flux in units of $10^{-21}~\rm{erg}~\rm{s}^{-1}~\rm{cm}^{-2}~\rm{Hz}^{-1}~\rm{sr}^{-1}$ for the case of Eddington-limited accretion at $1~\rm{kpc}$ distance from the BH. We consider initial BH masses of $100$ and $500~\rm{M}_{\odot}$, roughly the range expected for the direct collapse of massive Pop~III stars \citep{hw02,heger03}. As Fig.~14 shows, $J_{\rm{LW}}$ can greatly exceed the critical value of $\ga 10^{-2}$ required for the suppression of star formation in minihaloes, which mostly relies on efficient H$_{2}$ cooling \citep{mba01,yoshida03a,on08,jgb07b}. Furthermore, even a modest LW flux can dissociate enough H$_{2}$ such that HD formation and cooling never become important, reducing the temperature to which the gas can cool \citep{yoh07}. The impact on primordial star formation in BH-hosting galaxies might thus be severe even for sub-Eddington accretion rates, implying that this effect must be taken into account in future work.

Another important issue concerns the contribution to the global LW background. We may estimate a maximum global LW background by assuming that each atomic cooling halo at $z\ga 10$ hosts a BH accreting at the Eddington rate. We can then find an upper limit to the global LW background by summing up the contributions from BHs within a distance equal to the maximum mean free path for a LW photon, $r_{\rm{max}}\sim 10~\rm{Mpc}$ at $z=15$, following the prescription in \citet{jgb07b}. This estimate yields a maximum global LW background comparable to the LW flux from a single source, as shown in Fig.~14. Such a high flux could have profound consequences for further star formation in minihaloes, but in most cases radiative feedback by the progenitor star will significantly delay accretion, such that a global LW background fuelled by accretion on to BHs will likely be subdominant compared to a stellar LW background \citep{pdc07}.

\subsubsection{Ionizing flux}
The amount of ionizing radiation released by the accreting BH can be determined in analogy to our calculation of the LW flux. An integration over the temperature profile of the accretion disc yields the total number of hydrogen-ionizing photons emitted per second, shown in the bottom panel of Fig.~14. While massive Pop~III stars emit of the order of $10^{50}$ hydrogen-ionizing photons per second, these stars live for only $\la 3~\rm{Myr}$ \citep{bkl01,schaerer02}. However, as Fig.~14 shows, if Pop~III relic BHs are able to accrete efficiently, they may emit $10$~--~$100$ times this number for $\ga 100~\rm{Myr}$. This enormous flux of ionizing radiation could power H~{\sc ii} regions with radii of the order of $10~\rm{kpc}$, larger and longer-lived than the transient H~{\sc ii} regions of individual Pop~III stars \citep{yoshida07,jgb07a}. Star formation in minihaloes within the H~{\sc ii} region could be suppressed if accretion is continuous and drives a persistent radiative flux \citep{as07,whalen07}. Due to this dramatic radiative feedback associated with high accretion rates, though, the gas in the protogalaxy is likely to be heated and driven away from the BH, once again resulting in sub-Eddington accretion rates \citep{pdc07}.

We have also calculated the number of He~{\sc ii}-ionizing photons emitted by the accreting BH, and find that this is within a factor of $\la 2$ of the number of hydrogen-ionizing photons, owing to the high temperatures of the accretion disc. Thus, if Pop~III relic BHs are able to accrete efficiently, they may also contribute to the reionization of helium, driving He~{\sc iii} regions that can be as large as their H~{\sc ii} regions \citep{fo07}.

\begin{figure}
\begin{center}
\includegraphics[width=8cm]{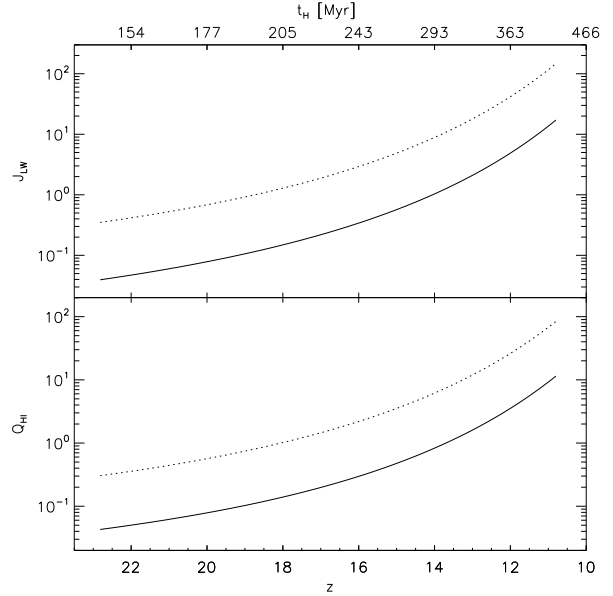}
\caption{The radiation due to thermal emission from the accretion disc surrounding a Pop~III relic BH accreting at the Eddington limit, for an inital mass of $100$ (solid lines) and $500~\rm{M}_{\odot}$ (dotted lines). {\it Top panel:} The photodissociating flux, $J_{\rm{LW}}$, in units of $10^{-21}~\rm{erg}~\rm{s}^{-1}~\rm{cm}^{-2}~\rm{Hz}^{-1}~\rm{sr}^{-1}$, at a distance of $1~\rm{kpc}$ from the BH. {\it Bottom panel:} The number of hydrogen-ionizing photons emitted per second from the accretion disc, in units of $10^{50}~\rm{s}^{-1}$.}
\end{center}
\end{figure}

\section{Summary and conclusions}
We have investigated the formation of the first galaxies with highly resolved numerical simulations, taking into account all relevant primordial chemistry and cooling. The first galaxies form at redshifts $z\ga 10$ and are characterized by the onset of atomic hydrogen cooling, once the virial temperature exceeds $\simeq 10^{4}~\rm{K}$, and their ability to retain photoheated gas. We have described the merger history of a $\simeq 5\times 10^7~\rm{M}_{\odot}$ system in great detail and found that in the absence of stellar feedback $10$ Pop~III.1 stars form in minihaloes prior to the assembly of the galaxy. Infalling gas is partially ionized at the virial shock and forms a high amount of H$_{2}$ and HD, allowing the gas to cool to the temperature of the CMB and likely form Pop~III.2 stars with $\ga 10~\rm{M}_{\odot}$. Accretion on to the galaxy proceeds initially via hot accretion, where gas is accreted directly from the IGM and shock-heated to the virial temperature, but is quickly accompanied by a phase of cold accretion, where the gas cools in filaments before flowing into the parent halo with high velocities. The latter drives supersonic turbulence at the centre of the galaxy and thus plays a key role for second-generation star formation. Finally, we have investigated the growth of BHs seeded by the stellar remnants of Pop~III.1 stars and found that accretion at the Eddington limit might be possible under the most favourable circumstances, but in most cases radiation emitted by the progenitor star and the accretion disc around the BH will lead to sub-Eddington accretion rates.

Depending on the strength of radiative and SN-driven feedback, some galaxies might remain metal-free and form intermediate-mass Pop~III.2 stars. The inclusion of radiative feedback would likely increase the fraction of Pop~III.2 material, as it enhances the degree of ionization, and, consequently, the amount of molecules formed \citep{jgb07a,wa07c}. Observational signatures of intermediate-mass primordial stars might include gamma-ray bursts (GRBs) for the case of a rapidly rotating progenitor \citep[e.g.][]{bl06}, or distinct abundance patterns produced by core-collapse SNe experiencing fallback \citep{un03}. However, since the number of Pop~III.1 stars formed in minihaloes prior to the assembly of the galaxy is of the order of $10$, it seems much more likely that at least one star will end in a violent SN explosion and pre-enrich the halo to supercritical levels \citep{byh03,greif07,wa07c}. In combination with the onset of turbulence, metal mixing in the first galaxies will likely be highly efficient and could lead to the formation of the first low-mass star clusters \citep{cgk08}, in extreme cases possibly even to metal-poor globular clusters \citep{bc02}. Some of the extremely iron-deficient, but carbon and oxygen-enhanced stars observed in the halo of the Milky Way may thus have formed as early as redshift $z\simeq 10$.

In future work, we plan to include the effects of radiative and SN-driven feedback by previous star formation in minihaloes, as well as the distribution of metals and its concomitant cooling. We will study the fragmentation of gas accumulating at the centre of the galaxy and address the issue of turbulence-driven star formation in detail. The goal of making realistic predictions for the first generation of starburst-galaxies, to be observed with the {\it James Webb Space Telescope (JWST)}, is clearly coming within reach.

\section*{Acknowledgments}
T.G. would like to thank Chris Burns at the Texas Visualization Laboratory for help with the images presented in this paper. V.B. acknowledges support from NSF grant AST-0708795 and NASA {\it Swift} grant NNX07AJ636. The simulations presented here were carried out at the Texas Advanced Computing Centre (TACC).

\end{document}